\documentstyle[aps,prd,eqsecnum,twocolumn,epsf]{revtex}

\newcommand{\beq}{\begin{equation}}
\newcommand{\beqn}{\begin{eqnarray}}
\newcommand{\eeq}{\end{equation}}
\newcommand{\eeqn}{\end{eqnarray}}
\newcommand{\beqa}{\begin{eqnarray}}
\newcommand{\eeqa}{\end{eqnarray}}

\newcommand{\vp}{\varphi}

\newcommand{\gsim}{\mbox{\raisebox{-1.ex}{$\stackrel
     {\textstyle>}{\textstyle\sim}$}}}
\newcommand{\lsim}{\mbox{\raisebox{-1.ex}{$\stackrel
     {\textstyle<}{\textstyle \sim}$}}}
\newcommand{\square}{\kern1pt\vbox{\hrule height
1.2pt\hbox{\vrule width 1.2pt\hskip 3pt
   \vbox{\vskip 6pt}\hskip 3pt\vrule width 0.6pt}\hrule
height 0.6pt}\kern1pt}

\def\beq{\begin{equation}}

\def\eeq{\end{equation}}

\begin{document}

\draft
\twocolumn[\hsize\textwidth\columnwidth\hsize\csname
@twocolumnfalse\endcsname

\title{{\bf When can preheating affect the CMB?}}
\author{Shinji Tsujikawa$^*$ and Bruce A.  Bassett$^{\P}$}
\address{$^*$ Research Center for the Early Universe,
University of Tokyo, Hongo,
Bunkyo-ku, Tokyo 113-0033, Japan \\[.3em]}
\address{$^{\P}$ Institute of
Cosmology and Gravitation, University of Portsmouth, Mercantile House,
Portsmouth PO1 2EG, \\ United Kingdom \\[.3em]}
\date{\today}
\maketitle
\begin{abstract}
We discuss the principles governing the selection of inflationary
models for which preheating can affect the CMB. This is a (fairly
small) subset of those models which have non-negligible
entropy/isocurvature perturbations on large scales during
inflation. We study new models which belong to this class -
two-field inflation with  negative nonminimal coupling and
hybrid/double/supernatural inflation models where the tachyonic
growth of entropy perturbations can lead to the variation of the curvature 
perturbation, ${\cal R}$, on super-Hubble scales.  
Finally we present evidence against recent claims 
for the variation of ${\cal R}$ in the 
absence of substantial super-Hubble entropy perturbations.
\end{abstract}
\vskip 2pc
]

\section{Introduction}                           %

The cosmological landscape is now dominated by a myriad of
inflationary models \cite{inf}, each with slightly different
genetics but a common origin, forged in the heyday of Grand
Unified field theories \cite{alan}.  Inflationary models are
flexible, hardy and spawn \& multiply with remarkable facility -
features flowing from their scalar field DNA.

And while the recent M-theory and brane-world revolution has again
stimulated interest in alternatives to inflation \cite{eky} it
remains true that the simplest inflationary models provide a very
good fit to current cosmological observations \cite{Boom01}.
Indeed, the Cosmic Microwave Background (CMB) and large scale
structure observations presently show no particular signature -
such as primordial non-Gaussianity - which might allow us to
single out a particular model or falsify inflation as a paradigm.
It is therefore good science to seek new ways to select the
``fittest" of inflationary models.

Preheating after inflation may provide exactly such a selection
rule. Preheating is probably the most violent of putative phases
in cosmic history \cite{TB,KLS}.  It requires non-perturbative
quantum field theory pushed to its non-equilibrium limits
\cite{Boy}.  The exponentially rapid population of sub- and
super-Hubble wavelength modes makes it possible to produce very
massive particles in large quantities and hence over-produce
dangerous relics such as modulini or gravitini \cite{GM}. In
addition there is some (admittedly speculative) evidence that the
induced parametric growth of small-wavelength metric perturbations
around the Hubble scale may lead generically to runaway production
of primordial black holes (PHB) \cite{PBH}.  If correct this will
rule out wide regions of the preheating parameter space.

At the other end of the cosmic scale, a natural question is
whether preheating can affect the modes of the metric
perturbations which induce the intricate anisotropies in the CMB,
and if so, what are the criteria? In this paper we will discuss
the conditions where preheating can lead to the growth of metric
perturbations on cosmic scales.

The answer to this question appears to be intimately linked to the
issue of correlations between adiabatic and entropy perturbations
from multi-field inflation.

\section{General principles}

In this section we present the conditions  known to be 
{\em sufficient} for preheating to affect the CMB; distilled from the
recent literature \cite{metpre}-\cite{Zibin}. Perhaps the best way
to discuss the impact on the CMB is via quantities which are
time-independent in the presence of only adiabatic perturbations
on large scales.  Such quantities are used to normalize the inflationary
models to the large-angle COBE measurements of the CMB.

Two common choices for such quantities are $\zeta$, the curvature
perturbation in the constant density hypersurfaces ($\delta \rho =
0$), and the comoving curvature perturbation ${\cal R}$. These two
coincide in the $k \to 0$ limit \cite{GWBM} (modulo a minus sign)
and we will use ${\cal R}$.

The existence of entropy perturbations is crucial for preheating
to affect the CMB, since in general \cite{multi}
\beq
\dot{{\cal R}} \rightarrow 3H\frac{\dot{p}}{\dot{\rho}} {\cal S}\,,
\label{zetagen}
\eeq
where the above limit is understood as $k \rightarrow 0$; $H$,
$p,\rho$ are the Hubble rate, the total pressure and energy
density respectively and ${\cal S}$ is the total (dimensionless)
entropy perturbation defined by
\beq
{\cal S} = H\left(\frac{\delta p}{\dot{p}} - \frac{\delta
\rho}{\dot{\rho}}\right)\,.
\label{entropygen}
\eeq

Now in the case of two minimally coupled scalar fields $\varphi_1$ and
$\varphi_2$, this can be given explicitly \cite{GWBM} as
\beq
\dot{{\cal R}} \rightarrow \frac{2H}{\dot{\sigma}} \dot{\theta}
\delta s.
\label{zetaspec}
\eeq
where $\sigma$ and $s$ are ``adiabatic'' and ``entropy''  fields,
defined by
\beqn
d\sigma &=& (\cos \theta)d\vp_1 +(\sin \theta)d\vp_2, \\
ds &=& -(\sin \theta) d\vp_1 +(\cos \theta) d\vp_2\,.
\label{sigs}
\eeqn
Here $\theta$ is the angle of the trajectory in $(\phi,\chi)$
field space, satisfying $\tan
\theta=\dot{\varphi}_2/\dot{\varphi}_1$. Eq.~(\ref{zetaspec})
means that variation of ${\cal R}$ requires {\em not only} a large
scale entropy perturbation $\delta s$, but also a non-straight
trajectory in field space. In the minimally coupled two-field
system with an effective potential $V(\varphi_1,\varphi_2)$, the
Fourier modes for the adiabatic and entropy field perturbations
satisfy \cite{GWBM}
\beqn
& &\delta \ddot{\sigma}+3H\delta \dot{\sigma}+\left( \frac{k^2}{a^2}
+V_{\sigma \sigma}-\dot{\theta}^2 \right)\delta \sigma \nonumber \\
&=& -2V_{\sigma}\Phi+4\dot{\sigma}\dot{\Phi}+
2(\dot{\theta}\delta s)^{\cdot}-\frac{2V_{\sigma}}
{\dot{\sigma}} \dot{\theta} \delta s,
\label{deltasig}
\eeqn
\beqn
\delta \ddot{s}+3H\delta \dot{s}+\left( \frac{k^2}{a^2}
+V_{ss}+3\dot{\theta}^2 \right)\delta s
= \frac{\dot{\theta}}{\dot{\sigma}}
\frac{k^2}{2\pi G a^2}\Phi,
\label{deltas}
\eeqn
where $k$ is a comoving momentum, $a$ is a scale factor,
$G$ is a Newton's gravitational constant, and
\beqn
V_{\sigma \sigma} &\equiv& (\cos^2 \theta) V_{\vp_1 \vp_1} +(\sin
2\theta)V_{\vp_1 \vp_2}+(\sin^2 \theta) V_{\vp_2 \vp_2},\\
V_{ss} &\equiv& (\sin^2 \theta) V_{\vp_1 \vp_1} -(\sin 2\theta)V_{\vp_1
\vp_2}+(\cos^2 \theta) V_{\vp_2 \vp_2}.
\label{Vdd}
\eeqn
$\Phi$ is a gravitational potential in the longitudinal gauge, satisfying
\beqn
\dot{\Phi}+H\Phi=4\pi G \dot{\sigma} \delta\sigma.
\label{Phisource}
\eeqn
Eq.~(\ref{Phisource}) indicates that the gravitational potential is
sourced by adiabatic field perturbations.

When the effective mass of $\delta s$ is light relative to the
Hubble rate $H=\dot{a}/a$ during inflation, i.e.,
\beqn
\mu_s^2 \equiv V_{ss}+3\dot{\theta}^2~\lsim~H^2\,,
\label{sup}
\eeqn
the entropy field perturbation is {\it not} suppressed on
super-Hubble scales during inflation. Then during preheating if
$\delta s$ is resonantly amplified due to a time-dependent
effective mass, this can lead to the growth of ${\cal R}$ on large
scales, thereby altering the power spectrum normalization or even
leaving the model incompatible with the large-angle CMB. Note that
the adiabatic field perturbation is sourced by the entropy field
perturbation, thereby stimulating the growth of $\Phi$ through
Eq.~(\ref{Phisource}). In contrast, if the entropy perturbation is
heavy during inflation, ($\mu_{s}^2 \gg H^2 $) then $|\delta s|
\sim a^{-3/2}$ and the growth during preheating means that the
change of ${\cal R}$ is negligible before backreaction ends the
resonance.

In Sec.~\ref{cmbaffected} we present new classes of models which
have strong preheating or tachyonic growth but simultaneously have
a light entropy perturbation in the preceding inflationary phase.
To begin with we shall analyze the evolution of ${\cal R}$ in the
massive chaotic inflationary scenario in the next section.

\section{An evaluation of claims for varying ${\cal R}$ in the absence of entropy perturbations}
There have been recent claims by Henriques and Moorhouse (HM)
\cite{HM} that ${\cal R}$ or $\zeta$ will vary during reheating or preheating even in the
{\em absence of large-scale entropy perturbations} when going beyond linear
perturbation theory and taking into account the quantum-to-classical
transition.

If correct this would have a profound impact on inflationary
cosmology \cite{mp2,mp3,mp4}. Our aim in this section is to
evaluate these claims. To do so let us consider metric preheating
in the massive chaotic inflationary scenario with a standard
four-leg interaction:
\beqn
V=\frac12 m^2\phi^2+\frac12 g^2\phi^2\chi^2\,.
\label{massive}
\eeqn
Strong amplification of the $\chi$ fluctuation  requires a
resonance parameter $q=g^2\phi^2/(4m^2) \gg 1$ at the beginning of
preheating \cite{KLS}.  In this case the field $\chi$ is heavy
during inflation relative to the Hubble rate \cite{suppression},
which means that $\chi$ is strongly suppressed during inflation
($\chi \sim a^{-3/2}$), thereby leading to $\dot{\theta} \simeq 0$
and $\langle V_{ss}\rangle \simeq \langle V_{\chi \chi}\rangle
\simeq g^2\phi^2 \gg H^2$. Therefore large scale entropy
perturbations are exponentially suppressed during inflation, which
safeguards nonadiabatic growth of super-Hubble curvature
perturbations, as found by Eq.~(\ref{zetaspec}).
Hence, at the linear level, $\dot{\cal R} \simeq 0$ to high precision in this 
model.

\begin{figure}
\epsfxsize = 3.8in
\epsffile{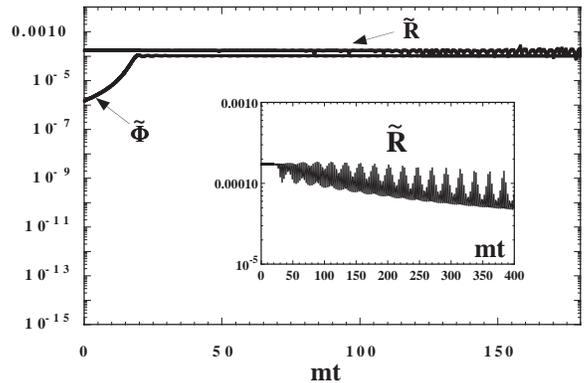}
\caption{The evolution of a super-Hubble curvature perturbation $\tilde{\cal R}
\equiv k^{3/2}{\cal R}$ and the gravitational potential $\tilde{\Phi} \equiv
k^{3/2} \Phi$ during inflation and preheating in the model (\ref{massive})
with $g=5.0 \times 10^{-4}$.  The initial conditions are chosen to be
$\phi=3M_{\rm pl}$ and $\chi=10^{-3}M_{\rm pl}$, in which case inflation
ends around $mt \sim 20$.
{\bf Inset}: The plot of $\tilde{\cal R}$ for large initial variance,
$\langle \delta \phi^2\rangle = 10^{-4}\phi_0^2$.}
\label{mchaotic}
\end{figure}

Nevertheless we have to caution that $\dot{\cal R}$ in
Eq.~(\ref{zetaspec}) is the result of the first order perturbation
theory.  If we include second order backreaction effects in the
background evolution equations, these give rise to second order
terms such as $\langle \delta \dot{\phi}^2 \rangle$ in the time
derivative of $\dot{\cal R}$ \cite{HM}.  Since these variances are
not suppressed during inflation (due to the short-wavelength
contributions), they may provide additional source terms for the
curvature perturbation. This, in essence, is the origin of the
claims by HM.

To check these claims we performed  our numerical simulations
implementing the second order backreaction effects as spatial
averages in the evolution equations \cite{KLS,Boy,selfTBV,selfZBS}
and compared the evolution of the cosmological perturbations found
using different (but equivalent at linear order) equations of
motion.

In Fig.~\ref{mchaotic} we plot the evolution of a super-Hubble
curvature perturbation using its definition, i.e., ${\cal
R}=\Phi-H/\dot{H}(\dot{\Phi}+H\Phi)$. The variance $\langle
\delta\chi^2 \rangle$ grows by parametric resonance until
backreaction  becomes important, while the inflaton fluctuation is
not excited unless rescattering is taken into account \cite{KT}.

The coherent oscillation of the inflaton condensate is destroyed
once backreaction starts to dominate.  This can affect the
evolution of $H$, $\Phi$ through Eq.~(\ref{Phisource}) and also
${\cal R}$.  In fact the curvature perturbations exhibits small
oscillations, but it remains roughly constant (see
Fig.~\ref{mchaotic}). When we use the first order equation
(\ref{zetaspec}), we found that ${\cal R}$ is conserved without
oscillations.

This is in stark contrast to the simulations of HM \cite{HM}
who found that
${\cal R}$ varies after the energy density of the inflaton condensate drops
below its variance $\langle \delta \phi^2\rangle$.  According to their
numerical results using similar second order approximations such as ours,
the {\it decrease} of ${\cal R}$ occurs in the preheating stage.

A large difference between the two investigations is that the
initial energy density of the variance $\langle \delta
\phi^2\rangle$ is smaller than $\phi_0^2$ only by four orders of
magnitude in Ref.~\cite{HM}, while we argue that the typical size
is regulated to be $\langle \delta \phi^2\rangle \sim m^2 \sim
10^{-12}M_{\rm pl}^2$ \cite{KT}, which is by ten orders of
magnitude smaller than $\phi_0^2$.

Since $\langle \delta \phi^2\rangle$ does not exceed $\phi_0^2$
during preheating in our simulations, we do not find the decrease
of ${\cal R}$ claimed by \cite{HM}. Using the initial conditions
of HM we checked that curvature perturbations exhibit small
decrease after $\phi_0^2$ drops under $\langle \delta
\phi^2\rangle$ (see the inset of Fig.~1), but we still {\em do
not} find the extensive changes reported in \cite{HM}.

We did find that implementing the numerics is subtle and subject
to artificial instabilities. Further, not all definitions or
equations for ${\cal R}$ are equally suitable for numerical
implementation, some being more susceptible to these
instabilities. We therefore argue that ${\cal R}$ probably does
not evolve on large scales in the absence of entropy perturbations
even at second order. This is consistent with the standard view
based on causality \footnote{The standard view simply associates
the second order terms to radiation with a temperature $T \propto
\sqrt{\langle \delta\chi^2\rangle}$. In the absence of large scale
entropy perturbations in this radiation fluid no changes are
induced in ${\cal R}$.}.

We end with one caveat, however: full lattice simulations show
that the growth of $\delta\chi$ leads to the excitation of
inflaton fluctuations via rescattering, thereby satisfying the
condition $\langle \delta \phi^2\rangle~\gsim~\phi_0^2$ around the
end of preheating \cite{KT}. It would be worth investigating
further whether any change in ${\cal R}$ occurs in such a
situation, although if they do, they are likely to be small.

\section{Models with the CMB affected by preheating} \label{cmbaffected}

\subsection{Chaotic inflation with self-interaction and
nonminimal coupling}

Achieving a light entropy perturbation during inflation is {\em
typically} rather difficult in chaotic inflation models. However,
this picture is modified if one takes into account nonminimal
coupling \cite{selfBV}. Let us consider the quartic chaotic
inflationary scenario in the presence of a nonminimally coupled
scalar field $\chi$ coupled to $\phi$ with effective potential:
\beqn
V=\frac14 \lambda \phi^4+\frac12 g^2\phi^2\chi^2+\frac12 \xi R \chi^2\,.
\label{selfnon}
\eeqn
In this case the effective mass of $\chi$ is given by
\beqn
m_{\rm eff}^2 \equiv g^2\phi^2+\xi R \approx
\lambda \phi^2 \left[\frac{g^2}{\lambda}+8\pi \xi
\left( \frac{\phi}{M_{pl}}\right)^2 \right] \,,
\label{effmass}
\eeqn
where we used the approximation $R \approx 12H^2$ which assumes
$\chi \sim 0$ and is valid to zero order in the slow-roll
parameters. Since the $\xi R$ term decreases faster than the
$g^2\phi^2$ term, it is possible for $\chi$ to be light relative
to $H$   during inflation by allowing negative values of $\xi$.
When $\xi=0$ it was shown in
Refs.~\cite{selfBV,selfFB,selfTBV,selfZBS} that super-Hubble
cosmological perturbations probed by CMB experiments can be
amplified around the center of the first resonance band,
$g^2/\lambda=2$ (see also Ref.~\cite{GKLS}). For
$g^2/\lambda~\gsim~8$, using the Hartree approximation, the growth
of sub-Hubble field perturbations shuts off the resonance before
super-Hubble metric perturbations are enhanced
\cite{selfZBS,BPLT}.

\begin{figure}
\epsfxsize = 3.4in
\epsffile{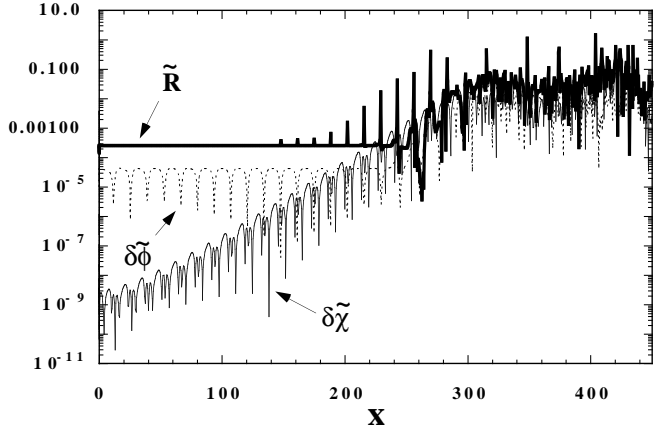}
\caption{The evolution of super-Hubble cosmological
 perturbations for $g^2/\lambda=18$ and $\xi=-0.12$
 with initial conditions $\phi=4M_{\rm pl}$ and $\chi=10^{-6}
 M_{\rm pl}$.  Nonminimal coupling makes the $\chi$ mass light, thereby leading
 to the amplification of ${\cal R}$ during preheating. Here $x$ is
 the  dimensionless conformal time.}
\label{self}
\end{figure}

However, this picture is modified by a negative nonminimal
coupling for $\chi$, which makes it possible to avoid the
inflationary suppression of the entropy perturbation even for
$g^2/\lambda~\gsim~8$. For example, when $g^2/\lambda=18$ and
$\xi=-0.12$ shown in Fig.~\ref{self}, the amplitude of the
super-Hubble $\delta\chi_k$ mode at the end of inflation is larger
than in the $\xi=0$ case by about 10 orders of magnitude.  In this
case, large-scale curvature perturbations exhibit nonadiabatic
growth after $\delta \chi$ catches up $\delta\phi$ (see
Fig.~\ref{self}).

\begin{figure}
\epsfxsize = 3.1in \epsffile{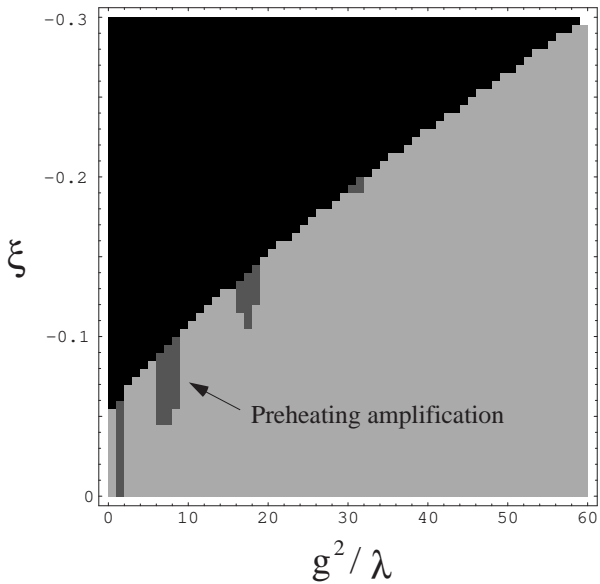} \caption{A density plot
in the ($g^2/\lambda, \xi$)-plane where the super-Hubble curvature
perturbations ${\cal R}$ is enhanced or not.  The black color and
the dark gray colours correspond to ranges where ${\cal R}$ grows
during inflation, and during preheating, respectively.  The region
where ${\cal R}$ does not vary is shown by the light gray color.
Note that there exist parameter ranges around the center of
resonance bands ($g^2/\lambda=2n^2$) where ${\cal R}$ grows during
{\it preheating} even for $g^2/\lambda \ge 8$. However, these
regions do not persist for $g^2/\lambda > 40$ as explained in the
text. } \label{denplot}
\end{figure}

When $g^2/\lambda \gg 1$, rather large negative nonminimal
coupling ($\xi~\lsim~-1$) is required to make the $\chi$ mass light.
In this case, if the $\xi R$ term gives almost the same contribution as
$g^2\phi^2$ at the beginning of inflation, it is impossible to avoid the
suppression of $\chi$ due to the decrease of $\xi R$ compared
to $g^2\phi^2$.
When $g^2\phi^2 <-\xi R$ initially, strong amplification of
${\cal R}$ occurs from the beginning of inflation \cite{TY,STY}.  Therefore
when $g^2/\lambda \gg 1$ with large negative nonminimal coupling, the
growth of ${\cal R}$ during inflation is significant rather than
during preheating.  The density plot of the parameter range where curvature
perturbations grow during inflation and preheating is shown in
Fig.~\ref{denplot}.

In the case of the massive inflaton and a $g^2\phi^2 \chi^2$ coupling,
the effective mass of $\chi$ is
approximately given by $m_{\rm eff}^2 \approx [g^2+16\pi
\xi (m/M_{pl})^2]\phi^2$ during inflation.
This implies that it is not possible to achieve both
the requirement of a light $\chi$ mass during inflation and a large $\chi$
mass ($q \gg 1$) after inflation and hence  ${\cal R}$ is not
enhanced during {\em preheating}
even in the presence of nonminimal coupling.

This situation changes, however, if we consider a cubic coupling of the
form $\tilde{g}^2  \phi \chi^2$ instead. This still leads to preheating
but now we can make $m_{\rm eff}^2$ light during inflation in certain regions
of the $(\tilde{g}/m, \xi)$ parameter space, analagous to those in
Fig.~\ref{denplot}.

\subsection{Hybrid and Double inflation}

In hybrid and supernatural inflationary
models \cite{hybrid1,hybrid2,RSG}, a symmetry breaking transition occurs
in the presence of a second scalar field, $N$. This leads to a
negative coupling/tachyonic instability in which the
longest wavelengths grow the strongest. Hence, even in the case
when the spectrum of  $N$ fluctuations is blue,
this long-wavelength instability may lead to variation of ${\cal R}$
in certain cases.

To investigate this in detail, consider Linde's hybrid inflation
potential \cite{hybrid1} with two scalar fields
$\phi$ and $N$:
\begin{eqnarray}
V= \frac{\lambda}{4} \left(N^2-\frac{M^2}{\lambda}\right)^2
+\frac12 g^2 N^2 \phi^2+\frac12 m^2\phi^2\,.
\label{hybrid}
\end{eqnarray}
Note that the supersymmetric version of the hybrid inflation
corresponds to the case, $g^2/\lambda=2$ \cite{hybrid2}.  Hereafter we shall
analyze the case where $g^2$ and $\lambda$ are the same order.

Inflation takes place due to the  slow-roll evolution of $\phi$
before the field reaches a bifurcation point $\phi_c=M/g$. If the
condition $m^2\phi_c^2 \ll M^4/\lambda$ is satisfied, the Hubble
constant at $\phi=\phi_c$ is given by $H=H_0 \equiv
\sqrt{2\pi/(3\lambda)}M^2/M_{\rm pl}$. It is convenient to
normalize the masses of two fields $\phi$ and $N$ relative to
$H_0^2$ as
\begin{eqnarray}
\alpha \equiv \frac{m^2}{H_0^2}=
\frac{3\lambda m^2 M_{\rm pl}^2}{2\pi M^4},
~~~~
\beta \equiv \frac{M^2}{H_0^2}
=\frac{3\lambda M_{\rm pl}^2}{2\pi M^2}\,.
\label{albe}
\end{eqnarray}
$\alpha$ is required to be smaller than unity in order to lead to
sufficient inflation for $\phi>\phi_c$. The evolution of the
fluctuation $\delta N$ depends on the value of $\beta$ as we will
classify below.

\subsubsection{Light $N$ field (Double inflation)}

When the $N$ mass is light ($\beta~\lsim~1$), the field $N$ and
its large scale perturbation are free from  inflationary
suppression ($\phi>\phi_c$). In this case the coupling $\lambda$
is constrained to be small, $\lambda~\lsim~(M/M_{\rm pl})^2$. In
order to end inflation due to a rapid rolling of the field $N$
after the symmetry breaking, one requires the ``water-fall''
condition $M^3 \ll \lambda mM_{\rm pl}^2$ \cite{hybrid1}.  If we
adopt the COBE normalization, $gM^5/(\lambda^{3/2}m^2M_{\rm pl}^3)
\simeq 3.5 \times 10^{-5}$, which comes from the single-field
fluctuation $\delta \phi$, the water-fall condition is not
typically satisfied due to small $\lambda$.  In this case the
second stage of inflation occurs after the symmetry breaking
(double inflation).

\begin{figure}
\epsfxsize = 3.5in
\epsffile{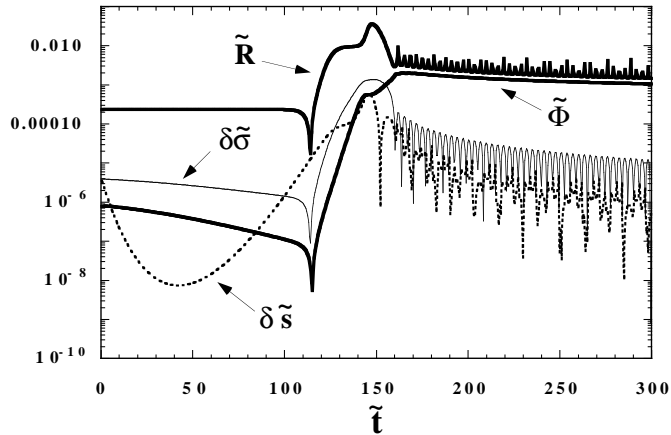}
 \caption{The evolution of super-Hubble cosmological
 perturbations in the case of a light $N$ mass with potential
 (\ref{hybrid}) as a function of $\tilde{t} \equiv \sqrt{2}Mt$
 using the Hartree approximation. We choose the initial conditions to be
 $\phi=1.5\phi_c$ and $N=10^{-3}M/\sqrt{\lambda}$.  We also show the
 evolution of adiabatic and entropy field perturbations, denoted by $\delta
 \sigma$ and $\delta s$, respectively.}
\label{lhybrid}
\end{figure}

Let us present a concrete example.  In Fig.~\ref{lhybrid} we plot
the evolution of ${\cal R}$ and $\Phi$ for $M=5.0 \times
10^{-6}M_{\rm pl}$, $g^2=\lambda=8.0 \times 10^{-11}$, and
$m=0.2M$, corresponding to mass parameters $\alpha=0.07$ and
$\beta=1.5$.  In this case the first stage of inflation
($\phi>\phi_c$) continues around $\tilde{t} \equiv \sqrt{2}Mt
\simeq 33$, after which the second phase of inflation takes place
until $\tilde{t} \simeq 130$.  The total number of e-folds is $N
\sim 70$ for the initial value, $\phi=1.5\phi_c$.

In Fig.~\ref{lhybrid} we find that the entropy perturbation
$\delta s$ is not strongly suppressed and begins to increase
after the symmetry breaking due to tachyonic growth of
the fluctuation $\delta N$.  This can lead to the amplification of
super-Hubble curvature perturbations until the backreaction of field
fluctuations shuts off its growth.

Note that we implement the field backreaction effect at second order in our
simulations.  Since the fluctuation $\delta N$ is efficiently amplified in
the tachyonic instability region, its growth terminates before the field
$N$ reaches the potential minimum.  The excitation of entropy perturbations
also leads to the amplification of adiabatic field perturbations $\delta
\sigma$ since they are correlated each other.  In Fig.~\ref{lhybrid} we
clearly find that the gravitational potential exhibits nonadiabatic growth
sourced by the growth of $\delta \sigma$.

During reheating no enhancement of super-Hubble cosmological
perturbations occurs in this case, since field perturbations are
already excited until the beginning of the oscillating phase. This
illustrates the importance of the spinodal instability region,
during which ${\cal R}$ and $\Phi$ are amplified as well as field
perturbations.

\subsubsection{Heavy $N$ field (Hybrid inflation)}

When $\beta \gg 1$, the field $N$ is exponentially suppressed
during inflation ($\phi>\phi_c$). In this case the water-fall
condition is typically satisfied after symmetry breaking, which
corresponds to the original hybrid inflationary scenario
\cite{hybrid1} where inflation ends due to the rapid rolling of
the field $N$.

The evolution of fluctuations for $\phi<\phi_c$ can be described
similarly as the models of spontaneous symmetry breaking analyzed
in Ref.~\cite{Felder}.  The field $N$ has practically no
homogeneous component at $\phi=\phi_c$.  In this case the
decomposition of $N$ between the homogenous field $N(t)$ and the
fluctuational part $\delta N({\bf x}, t)$ is not necessarily
valid.  Rather one is required to go beyond perturbation theory by
considering the full spatial distribution of the field $N({\bf x},
t)$.

\begin{figure}
\epsfxsize = 3.5in
\epsffile{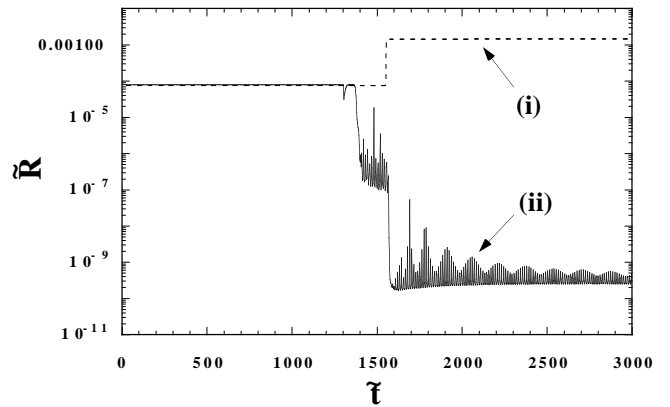}
 \caption{The evolution of the large scale curvature perturbation in the heavy
 $N$ case with $M=3.0 \times 10^{-6}M_{\rm pl}$, $g^2=\lambda=1.0 \times
 10^{-8}$ using the Hartree approximation to evaluate the backreaction.
 The result (i) is obtained by
 integrating Eq.~(\ref{zetaspec}), which is different from the result (ii)
 obtained by the definition ${\cal R}=\Phi-H/\dot{H}(\dot{\Phi}+H\Phi)$.
 In the case (ii) the field $\sigma$ is strongly dumped in
 Eq.~(\ref{Phisource}) due to overestimation of the field variance, thereby
 leading to the decrease of $\Phi$ and ${\cal R}$.
 In the case (i) the damping of $\sigma$ provides
 a source term for $\dot{\cal R}$.}
\label{hhybrid}
\end{figure}

If the perturbative approach based on the Hartree approximation is
applied, the effective mass of the field $N$ around $\phi=0$ is
approximately given by $m_{N}^2 \simeq 3\lambda \langle \delta N^2
\rangle-M^2$.  Naively one may expect that symmetry can be
restored due to the growth of $\delta N$ with $\langle \delta N^2
\rangle \simeq M^2/\lambda$, i.e., $m_N^2 \simeq 2M^2$.
Numerically we have found that the Hartree approximation leads to
symmetry restoration.  However we have to caution that this
approach neglects the rescattering effect and the formation of
topological defects. The authors in Ref.~\cite{Felder} found that
the field distribution reaches the potential minimum at
$N=M/\sqrt{\lambda}$ and $\phi=0$ instead of restoring symmetry
when using the 3-D lattice simulations without metric
perturbations. It was also pointed out that tachyonic growth of
the field fluctuation is typically followed by only a single
oscillation of the field.

In the Hartree approximation, the backreaction due to the growth
of long wavelength fluctuations prevents the field from reaching
the potential minimum due to an overestimation of the field
variance. In Fig.~\ref{hhybrid} we find that the evolution of
${\cal R}$ differs significantly when we integrate
Eq.~(\ref{zetaspec}) or evaluate ${\cal R}$ in terms of $\Phi$.
This disagreement obviously shows the limitations of linearized
gravity using the Hartree approximation.  It is certainly of
interest to analyze the evolution of super-Hubble cosmological
perturbations in the heavy $N$ case including full nonlinear
effects along the lines in Ref.~\cite{PE,FK}.

 \section{Conclusion and Discussion}

We discussed general principles governing the conditions under
which the large scale curvature perturbation, ${\cal R}$ - used to
normalize models to the large angle CMB anisotropies - is
amplified after (and during) inflation.

This is closely allied with the evolution of entropy/isocurvature
field perturbations $\delta s$ during the  inflationary phase. If
the mass of $\delta s$ is light relative to the Hubble rate, it
sources the variation of the curvature perturbation when it is
amplified during preheating or in the tachyonic instability
region.

In the model with $V=\frac12m^2\phi^2+\frac12g^2\phi^2\chi^2$,
super-Hubble entropy ($\delta s$) modes are exponentially
suppressed during inflation for the coupling $g$ required to have
efficient preheating.  Therefore growth of $\delta s$ during
preheating is not sufficient to lead to the variation of ${\cal
R}$ on large scales. This situation is not altered in the model
$V=\frac12m^2\phi^2+\frac12g^2\phi^2\chi^2+\tilde{g}^2\phi^3\chi$
where the fields evolve along the straight line ($\dot{\theta}
\simeq 0$) during inflation \cite{GWBM}.

In the model with $V=\frac14 \lambda\phi^4+\frac12g^2\phi^2\chi^2+
\frac12 \xi R \chi^2$, we found that parameter ranges where large
scale curvature perturbations are amplified during preheating are
significantly wider for negative nonminimal couplings.

We also studied hybrid/double inflationary models where
tachyonic/spinodal instability of the symmetry breaking field $N$
leads to the growth of entropy field perturbations.  When the
field $N$ is light in the first stage of inflation, super-Hubble
curvature perturbations typically exhibit tachyonic growth after
the symmetry breaking. If the field $N$ is heavy, which is the
original version of the hybrid inflationary scenario, we find that
linearized gravity theory using the Hartree approximation shows
some limitation in the sense that the evolution of ${\cal R}$ is
different depending on which  definition of ${\cal R}$ is used.

In the case of a purely quantum field with no classical vacuum
expectation value (vev) - such as a fermion field or a scalar
field whose vev vanishes during inflation such as discussed in
Sec.  IV B - a natural question arises: ``what are the correct
perturbed Einstein field equations?" It would be very useful to
develop a formalism in which these quantum fields were treated
non-perturbatively and self-consistently.

Finally, does inclusion of rescattering leads to extra variation
of ${\cal R}$ or does causality protects ${\cal R}$ (and suitable
nonlinear generalizations) to all orders in perturbation theory
\cite{tt} in the absence of large-scale entropy perturbations?
These interesting issues are left to future work.

\section*{ACKOWLEDGMENTS}
We thank Robert H.  Brandenberger for detailed and insightful comments
on the draft and Takahiro Tanaka for useful discussions.
ST is grateful for kind
hospitality during his stay at the University of Portsmouth. BB receives
support from EPSRC grant GR/R16488/01;  ST thanks
for financial support from the JSPS (No. 04942).


\end{document}